# Effects of ambient pressure on the subharmonic response from encapsulated microbubbles

**Nima Mobadersany[1], Amit Katiyar[2], Kausik Sarkar[1,2]**

**Department of Mechanical Engineering, The George Washington University, Washington DC**

**Department of Mechanical Engineering, University of Delaware, Newark DE**

## Abstract

Subharmonic response from contrast microbubbles as a function of ambient overpressure is numerically investigated for subharmonic aided noninvasive estimation of local organ level blood pressure. Three different interfacial rheological models for the encapsulation is used with material parameters appropriate for a common lipid coated contrast agent Sonazoid. The subharmonic response is seen to either decrease, increase or vary nonmonotonically with increasing ambient pressure. Compared to a free microbubbles important differences arise due to the encapsulation. Specifically due to the enhanced damping due to encapsulation, the range of excitation over which subharmonic is seen is broader than that in free microbubbles. This results in different trends of subharmonic response at the same excitation frequency for different excitation pressures. The observed behaviors are explained by investigating subharmonic generation threshold and resonance frequency.

# I. Introduction

Gas microbubbles (diameter $< 10$ μ) coated with lipids, proteins, polymers and other surface active materials are an ideal candidate for enhancing the contrast of diagnostic ultrasound images (deJong, 1996; Goldberg *et al.*, 2001; Ferrara *et al.*, 2007; Nahire *et al.*, 2014a; Nahire *et al.*, 2014b; Paul *et al.*, 2014). These ultrasound contrast agents (UCA) are strongly nonlinear; under a strong enough excitation they generate second harmonic (response at frequency $2f$ when excited at $f$) and subharmonic (response at frequency $f/2$) signals giving rise to harmonic (deJong *et al.*, 1994; Chang *et al.*, 1995) and subharmonic (Shankar *et al.*, 1998; 1999; Shi *et al.*, 1999a; Shi *et al.*, 1999b; Forsberg *et al.*, 2000) imaging modalities with superior contrast-to-tissue ratio. Recently, Forsberg et al have proposed and implemented a novel technique of using ambient pressure dependent subharmonic (at half the excitation frequency) response of UCAs for noninvasively estimating the local organ level blood pressure (Shi *et al.*, 1999d; c; Leodore *et al.*, 2007; Halldorsdottir *et al.*, 2011; Dave *et al.*, 2012a; Dave *et al.*, 2012b). In a previous article (Katiyar *et al.*, 2011), we demonstrated that as per single bubble dynamics model, subharmonic response from free microbubbles may either increases, decrease or vary non-monotonically with increasing ambient pressure. Here, we extend the study to encapsulated microbubbles numerically simulating its dynamics with different models of encapsulations and explain the underlying physics.

Local organ-level pressure estimation can provide critical information for accurate diagnosis of cardio-vascular diseases such as portal hypertension (Itai and Matsui, 1997; Pieters *et al.*, 1997) or heart valve defects (Marino *et al.*, 1985; Kasimir *et al.*, 2004). Currently available techniques include insertion of manometer-tipped catheter which, though accurately measures absolute local pressure, suffers from complications associated with such invasive techniques. On the other hand the non-invasive Doppler ultrasound has proved unreliable (Strauss *et al.*, 1993; Reddy *et al.*, 2003). Bubbles, because of their high compressibility, offer promises as an effective pressure sensor. In the seventies, Fairbank and Scully attempted to utilize the shift of microbubble resonance frequency with ambient pressure for pressure measurement (Fairbank and Scully, 1977). However, lack of a monodisperse microbubble suspension prevented a sharp resonance frequency delineation. Subsequent effort to improve this idea has faced limitations due to intrinsic lack of sensitivity of the technique (Hok, 1981; Ishihara *et al.*, 1988; Bouakaz *et al.*, 1999) .

Shi et al (Shi *et al.*, 1999d) found that subharmonic response varies far more strongly with ambient pressure than fundamental or second harmonic response from contrast agent Levovist—9.9 dB decrease with 24.8 kPa increase in ambient pressure at 2 MHz and 0.39 MPa excitation. Later the same group obtained linear decrease with a number of contrast agents—10.1 dB for Optison (GE Healthcare, Princeton, NJ), 11.03 dB for Definity (Lantheus Imaging, N. Billerica, MA), 12.2 dB for PRC-1 (Zhifuxian, Xinqiao Hospital, the Third Military Medical University, Chongqing, China), and 13.3 dB for Sonazoid (GE Healthcare, Oslo, Norway) (Leodore *et al.*, 2007; Halldorsdottir *et al.*, 2011). They have already investigated application of this technique for monitoring of portal hypertension (Dave *et al.*, 2012a) and tracking pressure in LV (Dave *et al.*, 2012b) and interstitial fluids in tumors (Halldorsdottir *et al.*, 2014). Adam et al (Adam *et al.*, 2005) found an 8 dB reduction in subharmonic response from UCA Optison with 40–140 mm Hg increase in ambient pressure. Andersen and Jensen in their experiment with UCA Sono Vue (Bracco, Milano, Italy) found the ratio of the subharmonic to fundamental to be a better measure for pressure variation, which showed a linear decrease (roughly 0.4 dB/kPa) with pressure increase at an excitation of 0.5 MPa and 4 MHz (Andersen and Jensen, 2010). In contrast to all these observations that showed subharmonic decrease with ambient pressure, phospholipid coated bubbles (similar to BR14 or Sono Vue) showed 9.6 dB decrease in subharmonic for 180 mm Hg increase in ambient pressure at 400kPa excitation; they also found a 28.9 dB increase in subharmonic at a low excitation of 50kPa (Frinking *et al.*, 2010).

Note that all but one experimental investigations noted above found a decrease in subharmonic response with ambient pressure increase. Intuitively one would expect a decrease in bubble activity with ambient pressure increase which is expected to restrict bubble activities. However, in our previous article, we conclusively demonstrated that the well-established Rayleigh-Plesset equation predicts that subharmonic response from a microbubble can either increase or decrease with ambient pressure, the behavior depending on the ratio of the excitation frequency to bubble resonance frequency (Katiyar *et al.*, 2011). The phenomenon was explained in light of resonance phenomenon, in that the behavior changes from increasing to decreasing as one either approach subharmonic resonance peak or recedes from it. However, we restricted the study to free microbubble so that the underlying physics can be understood free of the uncertainties of encapsulation models. Free microbubbles are unstable due to high Laplace pressure and dissolves away in milliseconds (Sarkar *et al.*, 2009). The encapsulation stabilizes against gas diffusion (Katiyar and Sarkar, 2010; 2012), but also critically affects the acoustic behaviors (deJong *et al.*, 1992;

Chatterjee and Sarkar, 2003; Paul *et al.*, 2014) including subharmonic response and its excitation threshold (Katiyar and Sarkar, 2011; 2012).

Henceforth, we will refer to our earlier article (Katiyar *et al.*, 2011) on pressure dependent subharmonic response of free microbubbles as KS11. There we briefly considered the case of encapsulated agents by simulating it using the code BUBBLESIM that assumes a linear viscoelastic shell model. This was to show that an earlier study (Andersen and Jensen, 2009) using BUBBLESIM found only decreasing trend of subharmonic response with increasing ambient pressure, is because the authors restricted to single frequencies—2.06 MHz for Levovist 2.46 MHz for Sonazoid. When excitation frequency was varied, one gets diverse subharmonic responses depending on frequencies—monotonically decreasing, increasing or nonmonotonically varying—with increasing ambient pressure. However, there has not been any systematic investigation of the underlying physics of pressure dependent subharmonic response from encapsulated microbubbles.

As noted before, there have been many models developed for contrast agents staring from simple yet effective representation of the encapsulating shell by effective lumped parameters (deJong *et al.*, 1992). Church (Church, 1995) considered the first rigorous model as a layer of viscoelastic material. In 2003, arguing that for a few-molecule-thick encapsulation, such a finite-thickness models are inappropriate, we introduced an interfacial rheological approach of modeling encapsulation (Chatterjee and Sarkar, 2003). Over the years, we hierarchically developed the original Newtonian model to a viscoelastic (CEM) (Sarkar *et al.*, 2005) and an exponentially elastic model (EEM) (Paul *et al.*, 2010). Marmottant et al introduced an interesting interfacial rheological model which combined a viscoelastic interface that has a buckling as well as a rupture radius. This model was very successfully applied to explain many phenomena experimentally observed using ultrahigh framerate optical setup (Marmottant *et al.*, 2005). In the following, we consider three different models of encapsulation—1) due to Church, and modified by Hoff et al (CH), 2) EEM model, and 3) Marmottant model (MM) to a typical Sonazoid microbubble. In Section II, we provide the mathematical description and the numerical technique. Section III presents the results and IV summaries the findings.

## II. Mathematical Formulation and Numerical Solution

## A. Governing equations

The dynamics of an encapsulated microbubble is governed by the Rayleigh-Plesset type equation. A number of different models have been proposed to account for the forces arising at the encapsulation, some treating the encapsulation as a thin layer of finite thickness characterized by bulk viscosity and elasticity, others treating it as a rheologically complex interface with interfacial viscosity and elasticity. We have recently shown that in the limit of infinitesimal thickness, they can all be cast in a common form:

$$\rho\left(R\ddot{R} + \frac{3}{2}\dot{R}^2\right) = P_{G_0}\left(\frac{R_0}{R}\right)^{3\kappa}\left(1 - \frac{3\kappa\dot{R}}{c}\right) - \frac{2}{R}\gamma(R) - \frac{4\dot{R}}{R^2}\kappa^s(R) - 4\mu\frac{\dot{R}}{R} - p_0 + p_A\sin\omega t. \qquad (1)$$

$R$ is the time dependent bubble radius, $\dot{R}$ and $\ddot{R}$ are the first and the second order time derivatives of the bubble radius, $c = 1485\,\text{m/s}$ is the velocity of sound in the surrounding liquid representing finite compressibility, $\rho = 1000\,\text{kg/m}^3$ is the liquid density, $\mu = 0.001\,\text{Ns/m2}$ is the liquid viscosity, $R_0$ is the initial bubble radius, $P_{G_0}$ is the initial inside gas pressure, $p_0$ is the ambient pressure and $p_A(t)$ is the excitation pressure. Gas diffusion during oscillation is neglected. The inside gas pressure obeys a polytropic law with index $\kappa$. Since with oscillations at MHz frequency Peclet number $Pe = R_0^2\omega / D_g \gg 1$ ($D_g$ is the thermal diffusivity; for $C_3F_8$ in side Sonazoid $2.8\times10^{-6}\,\text{m}^2\text{/s}$), we assume an adiabatic behavior for the gas inside ($\kappa = 1.07$ for $C_3F_8$). The effective surface tension $\gamma(R)$ and the surface dilatational viscosity $\kappa^s(R)$ describe the interfacial rheology of the encapsulation. For a free bubble $\gamma(R) = \gamma_w$, surface tension at a pure air-water interface and $\kappa^s(R) = 0$.

## B. Interfacial rheology of the encapsulation

We use three different models for the interfacial rheology, i.e. $\gamma(R)$ and $\kappa^s(R)$:

### 1. Viscoelastic model with exponentially varying elasticity (EEM) (Paul et al., 2010)

$$\gamma(R) = \gamma_0 + E^s\beta \quad \text{and} \quad \kappa^s(R) = \kappa^s \text{ (constant)}, \qquad (2)$$

where $\gamma_0$ is the constant interfacial tension, $E^s = E_0^s \beta \exp(-\alpha^s \beta)$, $\beta = R^2 / R_E^2 - 1$ is the area fraction change. Enforcing a balance of pressure and zero effective surface tension for stability $R_0^0$ we obtain equilibrium radius

$$R_E = R_0^0 \left[ 1 + \left( \frac{1 - \sqrt{1 + 4\gamma_0 \alpha^s / E_0^s}}{2\alpha} \right) \right]^{-1/2}.$$

The expression for the resonance frequency ( $f_0^0$ with initial radius $R_0^0$ ) due to EEM is given as

$$f_0^0 = \frac{1}{2\pi R_0^0} \sqrt{\frac{1}{\rho} \left( 3\kappa p_0 + \frac{2E_0^s}{R_0^0} \left( \frac{\sqrt{1 + 4\gamma_0 \alpha^s / E_0^s}}{\alpha^s} \right) \left( 1 + 2\alpha^s - \sqrt{1 + 4\gamma_0 \alpha^s / E_0^s} \right) \right)}. \tag{3}$$

2. *Marmottant model (MM) (Marmottant et al., 2005):*

$$\gamma(R) = \begin{cases} 0 & \text{for } R \leq R_{buckling} \\ \chi \left( \dfrac{R^2}{R_{buckling}^2} - 1 \right) & \text{for } R_{buckling} \leq R \leq R_{rupture} \\ \gamma_w & \text{for } R \geq R_{rupture} \end{cases} \text{ and } \kappa^s(R) = \kappa^s \text{ (constant)}, \tag{4}$$

where $\chi$ (same as $E^s$ in (2) ) is the elastic modulus of the shell, $R_{buckling} = R_0 \left[ 1 + \gamma(R_0) / \chi \right]^{-1/2}$

and $R_{rupture} = R_{buckling} \left[ 1 + \gamma_\omega / \chi \right]^{1/2}$. Above $R_{rupture}$, the bubble is assumed to have a pure air-water interface and below $R_{buckling}$, it is in a buckled state where the effective interfacial tension is zero. As can be seen there are additional parameters in this model such as $R_{buckling}$. We assume that the bubble is initially in a buckled state $R_0^0 = R_{buckling}$. Note that further increasing overpressure decreases the radius keeping it in a buckled state. Due to the discontinuous nature of (4) near buckling radius an analytical expression for the resonance frequency was not available. It was computed numerically.

*3. Church-Hoff model (CH) (Church, 1995; Hoff et al., 2000)*

$$\gamma(R) = 6G_s d_{sh_0} \frac{R_0^{0^2}}{R^2}\left(1 - \frac{R_0^0}{R}\right) \qquad \text{and} \qquad \kappa^s(R) = 3\mu_s d_{sh_0} \frac{R_0^{0^2}}{R^2} \quad . \tag{5}$$

This model treats the encapsulation as a layer of finite thickness $d_{sh_0}$ containing a viscoelastic material with shear modulus $G_s$ and shear viscosity $\mu_s$ here cast in an interfacial rheological assuming small $d_{sh_0}$. The resonance frequency is

$$f_0^0 = \frac{1}{2\pi R_0^0}\sqrt{\frac{1}{\rho}\left(3\kappa p_0 + 12G_s\frac{d_{sh_0}}{R_0^0}\right)}. \tag{6}$$

## C. Static change in initial radius due to overpressure

As noted in KS11, the static overpressure changes the initial bubble radius. Its value $R_0$ is different from the value $R_0^0$ at atmospheric pressure. Stability against gas diffusion at atmospheric pressure requires $\gamma(R_0^0) = 0$. We assume that gas exchange during static pressure change is negligible. It can be justified by noting that the encapsulation effectively hinders gas permeation. The gas diffusion can significantly decrease the bubble radius and effectively change the size distribution. We have recently executed a theoretical investigation modelling diffusion and resulting change radius decrease while analysing experimentally measured attenuation through a contrast agent suspension under overpressure. The results are based on entire bubble distribution. Here, however we are only interested in describing response from a single bubble, and want to avoid the uncertainties of determining gas permeability of the encapsulation. Gas permeability decreases the radius of a microbubble of initial radius 3 μm (the size considered extensively in the article) by 2.5% under an overpressure of 200 mmHg. The decrease is smaller for smaller initial radii. At static condition, considering the equation (1) at atmospheric pressure and at a finite overpressure one obtains

$$\frac{2\gamma(R_0)}{R_0} + P_{atm} + P_{ov} = \left(\frac{2\gamma(R_0^0)}{R_0^0} + P_{atm}\right)\left(\frac{R_0^0}{R_0}\right)^{3k}, \tag{7}$$

Knowing the static bubble radius $R_0^0$ at atmospheric pressure (7) is solved to find the initial radius of the bubble $R_0$ at other ambient pressures (i.e. in presence of nonzero overpressures).

**Table 1.** Encapsulation parameters for Sonazoid contrast agent used in the numerical simulation

| | |
|---|---|
| EEM(Paul *et al.*, 2010) | $\gamma_0 = 0.019 N/m$ , $E_0^s = 0.55 N/m$ , $\alpha^s = 1.5, \kappa^s = 1.2 \times 10^{-8} Ns/m$ |
| Church-Hoff model(Church, 1995; Hoff *et al.*, 2000) | $G_s = 52 MPa$ , $\mu_s = 0.99 Ns/m^2$ , $d_{sh_0} = 4 nm$ |
| Marmottant model (Marmottant *et al.*, 2005) | $\gamma(R_0) = 0.0$ , $x = 0.53 N/m$ , $\kappa^s = 1.2 \times 10^{-8} Ns/m$ |

## D. **Acoustic response of an encapsulated bubble**

The bubble dynamics equation (1) is solved using MATLAB® (Mathworks Inc., Natick, MA) to compute bubble radius. The scattered acoustic pressure is computed as

$$P_s(r,t) = \rho \frac{R}{r}(R\ddot{R} + 2\dot{R}^2),\tag{8}$$

After the scattered response is computed, one performs an FFT to obtain the components at various frequencies, specifically the subharmonic component of ½ order, at $f/2$ . Forsberg et al found that among the UCAs they studied—Definity, Levovist, Optison, ZFX, and Sonazoid—Sonazoid performed the best for pressure estimation. Therefore, in this study we use is as the model UCA. It has a lipid coating and a $C_4F_{10}$ gas core. In Table 1, we list the model parameters for Sonazoid used in the numerical simulation for each of the three encapsulation models considered. Sonazoid bubbles have a size distribution with number average radius 1.6μm. However, not all size contribute equally to subharmonic response. Subharmonic response from microbubbles occurs only above a threshold excitation level; threshold level increases with decreasing radius. In our earlier publication, we showed that the representative radius of the bubble that are contributing to the subharmonic response in the neighbourhood of the experimentally observed threshold, 200-400kPa and at 2 and 3 MHz excitation frequency is about 3μm. Therefore, in this paper, we present our results for this radius and then briefly consider the number average radius values.

## III.   Results and Discussion

### A.   Subharmonic response variation with ambient pressure

In KS11, for a free microbubble, we explained the subharmonic response variation with ambient pressure using the idea of subharmonic resonance. Just like the fundamental response, we showed that the subharmonic response from a microbubble as a function of excitation frequency normalized by its resonance frequency ( $f / f_0$ ) has a peak—subharmonic resonance. The peak was found near a value of about 1.65. The resonance frequency $f_0$ of a microbubble increases with increasing ambient pressure e.g. as per equation (3). Therefore, when the ambient pressure is increased, in the subharmonic resonance curve one either approaches the peak (for $f / f_0$ far above 1.65)—subharmonic response increases—or one recedes away from the peak (for $f / f_0$ far below 1.65)—subharmonic decreases. The subharmonic resonance curve typically was shown to have undulations near the peak value, which explained the nonmonotonic behavior in the intermediate excitation frequencies.

In this subsection, we consider EEM for encapsulation rheology leaving other models to a later subsection. In Figure 1(a), we plot the subharmonic response from a Sonazoid microbubble of initial radius 3μm as a function of $f / f_0$ at atmospheric pressure for different excitation amplitudes. Note that in KS11, we just plotted this curve for single excitation amplitude 0.24 MPa. Here curves are different for different excitations. For lower excitations 0.34 MPa and 0.36 MPa, one sees curves similar to the one seen for free bubbles in KS11. But for higher excitations, we do not see the sharp decrease away from the peak in the low frequency range. In Figure 1(b), (c) and (d), we plot the subharmonic variation with ambient pressure increase for different pressure excitations, 0.34 MPa, 0.38 MPa and 0.44MPa. In Figure 1(b), excitation amplitude 0.34 MPa falls below the threshold of subharmonic generation at higher ambient pressures for $f / f_0^0$ =1.4, 1.7 and 1.9; therefore the curves are only plotted till one gets subharmonic. Here for the lower frequency ratios ( $f / f_0^0$ <1.7) one sees monotonic decrease, and for higher frequency ratios one sees monotonic increase. Note that $f_0^0$ = 1.76 MHz. In Figure 1(c) at 0.38 MPa, $f / f_0^0$ = 1.2 and 1.3 show monotonic decrease, $f / f_0^0$ = 1.4 and 1.6 nonmonotonic variation and frequency ratio 1.8 and above showed monotonic increase. In Figure 1(d) at 0.44 MPa, for the high excitation frequencies $f / f_0^0$ = 1.8 and 2.1, subharmonic decreases. For the lower frequencies $f / f_0^0$ = 1.2 and 1.4, we see the nonmonotonic variation. We do not see any monotonic decrease at this excitation even at lower frequency ratios. Therefore, unlike the cases considered in KS11, here the excitation amplitude plays a crucial role.

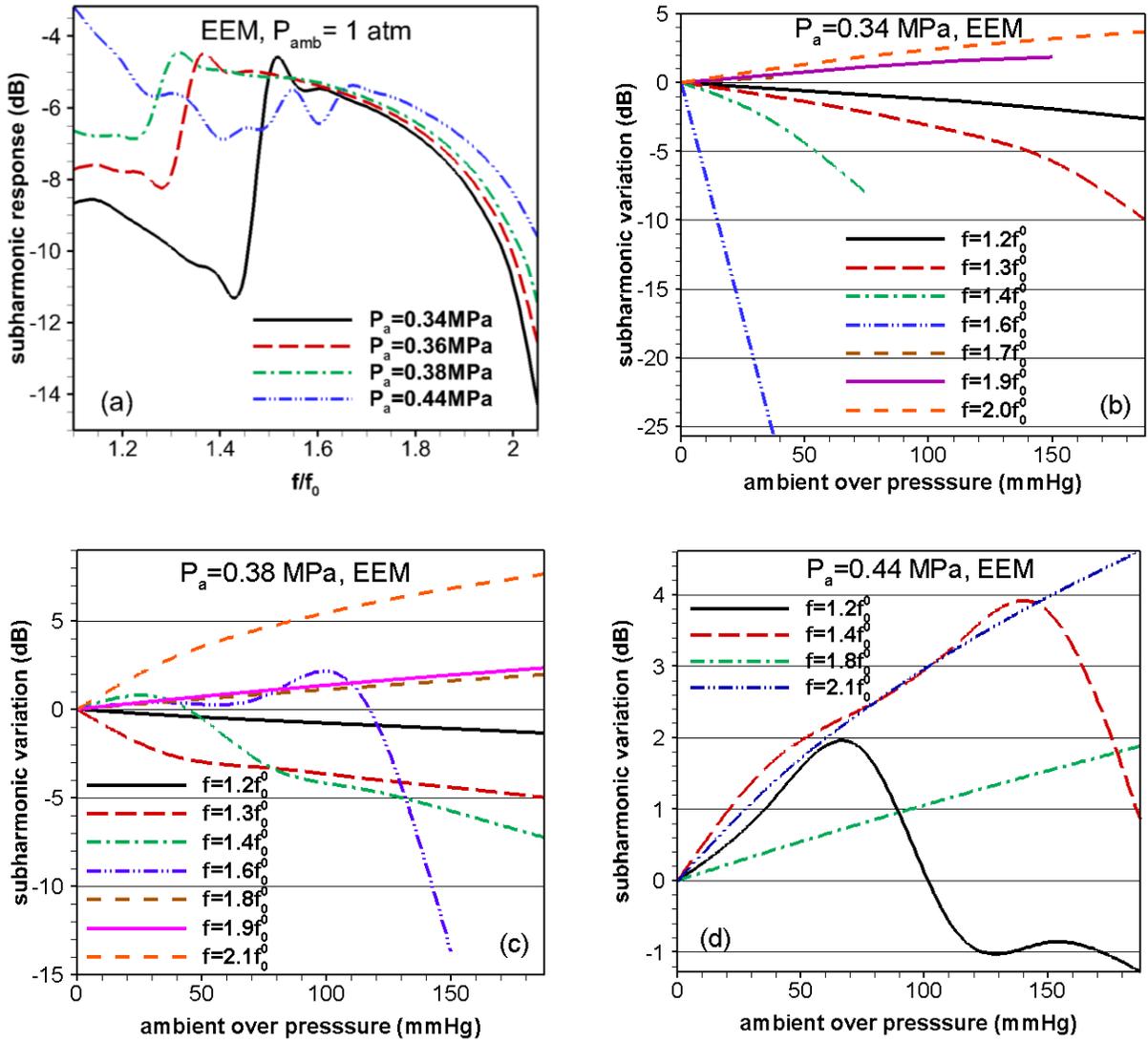

**Figure 1.** (a) Subharmonic response of Sonazoid microbubble ($R_0^0 = 3$ μm) as a function of $f / f_0^0$ ($f_0^0 = 1.76$ MHz). Subarmonic response as a function of ambient overpressure at excitation amplitude 0.34 MPa (b), 0.38 MPa (c) and 0.44 MPa (d) according to EEM model.

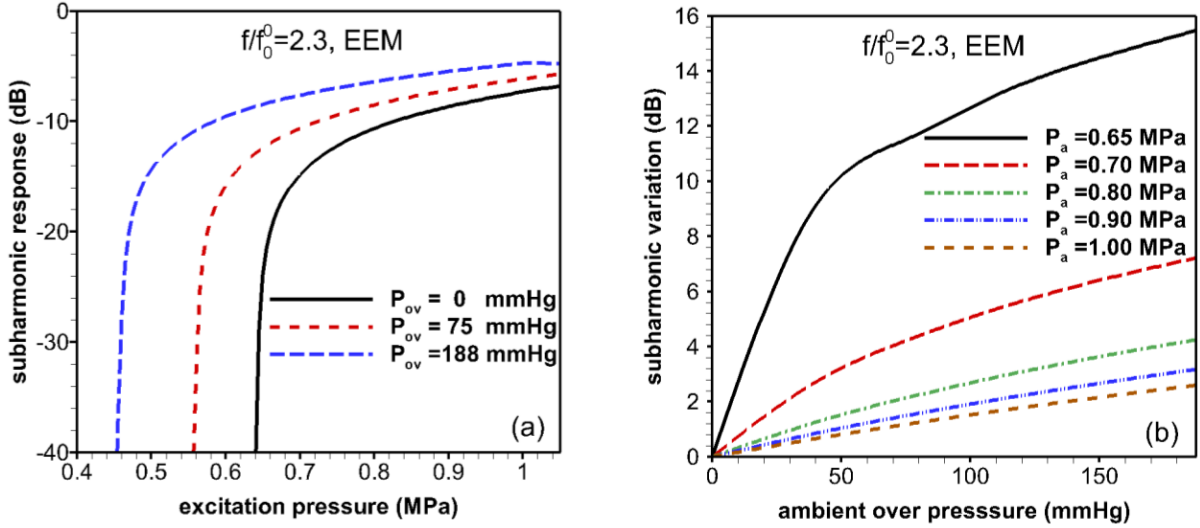

**Figure 2.** Subharmonic response of Sonazoid microbubble ($R_0^0 = 3$ μm) varying with excitation pressure (a) and with ambient over pressure (b) at $f / f_0^0 = 2.3$ according to EEM model.

Similar to KS11, we note here that at sufficiently high frequency ratios $f / f_0^0 \geq 2.0$, one does see a monotonic increase of subharmonic response with ambient pressure as long as the excitation is above threshold of subharmonic generation. It follows again from the sharp drop-off there in the subharmonic response versus frequency ratio (Figure 1a). We see it clearly in Figure 2 at $f / f_0^0 = 2.3$. Figure 2(a) plots the subharmonic response as a function of excitation pressure for several different ambient pressures. The threshold decreases with increasing ambient pressure. This generates the subharmonic variation with ambient pressures at different excitation pressures in Figure 2(b), where excitations are chosen above thresholds. Subharmonic response characteristically arises beyond a threshold excitation and then rapidly rises to reach a saturation level. The maximum sensitivity of subharmonic response with ambient pressure is achieved if the excitation level is carefully chosen close to the threshold, e.g. 0.65MPa, before the response reaches saturation.

At lower frequencies, subharmonic response is more complex than what we found in free microbubbles in KS11; often at the same frequency ratio $f / f_0^0$ one finds monotonic decrease or nonmonotonic variation depending on the excitation amplitude. In Figure 3, we plot the same quantities as in Figure 2 but at a lower frequency ratio $f / f_0^0 = 1.3$. Note that unlike Figure 2 (a) ($f / f_0^0 = 2.3$), here subharmonic threshold increases with increasing ambient pressure (Figure 3a), which gives rise to monotonically decreasing subharmonic response in Figure 3(b) for low excitation amplitudes 0.34 MPa and 0.37 MPa. For higher excitation levels the response becomes nonmonotonic because at those excitation levels, after reaching saturation subharmonic tends to precipitously fall and disappear (Figure 3a), e.g. at zero overpressure, there is no subharmonic beyond 0.45MPa.

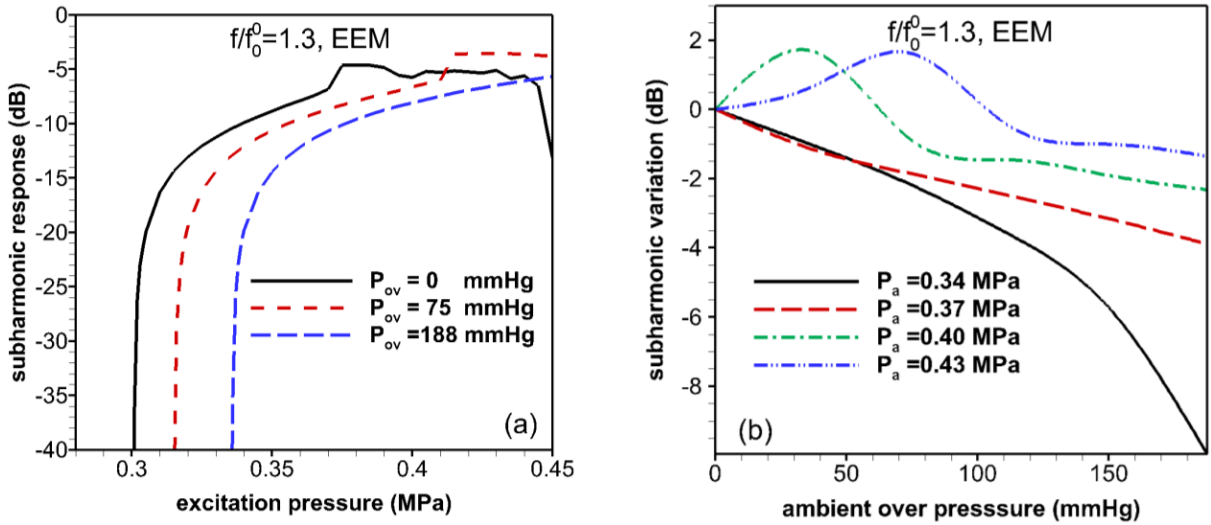

**Figure 3.** Subharmonic response of Sonazoid contrast agent ( $R_0^0 = 3$ μm) varying with excitation pressure (a) and with ambient over pressure (b) at $f / f_0^0 = 1.3$ according to EEM model.

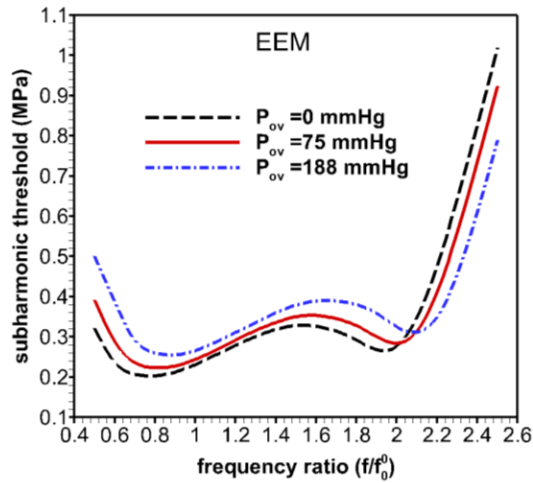

**Figure 4.** Subharmonic threshold of Sonazoid contrast agent ( $R_0^0 = 3$ μm) as a function frequency ratio according to EEM.

The reversal of trends in subharmonic thresholds with ambient overpressure in high and low frequency ratios is shown in Figure 4. For lower frequencies, the subharmonic threshold increases with ambient pressure, while for higher frequencies above $f / f_0^0 \approx 2.0$ the trend reverses. The crossing of the threshold curves can be explained by noting that resonance frequency $f_0$ increases over $f_0^0$ with ambient pressure increase. Therefore plotting the curve here as a function of $f / f_0^0$ instead of $f / f_0$ shifts the curves for higher ambient pressure to higher values of the abscissa, resulting in the crossing near $f / f_0^0 \approx 2.0$ when the curves sharply rise with increasing frequency. The increase of subharmonic response with increasing ambient pressure for larger frequency ratios therefore can be related to threshold being lower at higher ambient pressure. The lower thresholds for higher overpressures above $f / f_0^0 \approx 2.0$ allows subharmonic response to grow more giving rise to increasing subharmonic with increasing overpressure. For frequency ratios below the crossing, the opposite happens.

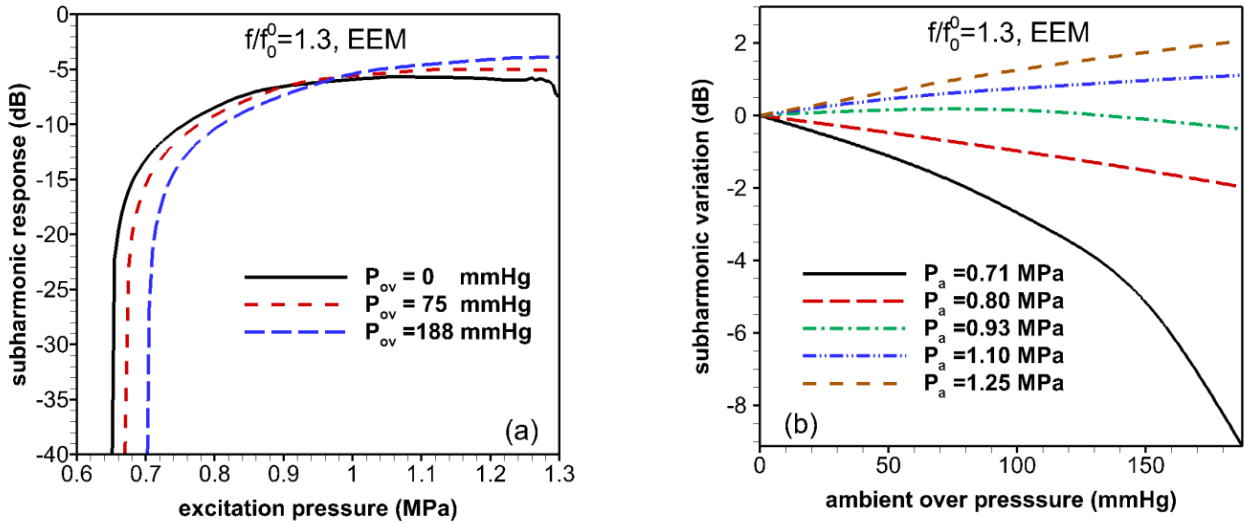

**Figure 5.** Subharmonic response of Sonazoid contrast agent varying with excitation pressure (a) and with ambient over pressure (b) at $f / f_0^0 = 1.3$ ($f_0^0 = 4.2$ MHz) with $R_0^0 = 1.6$ μm according to EEM model.

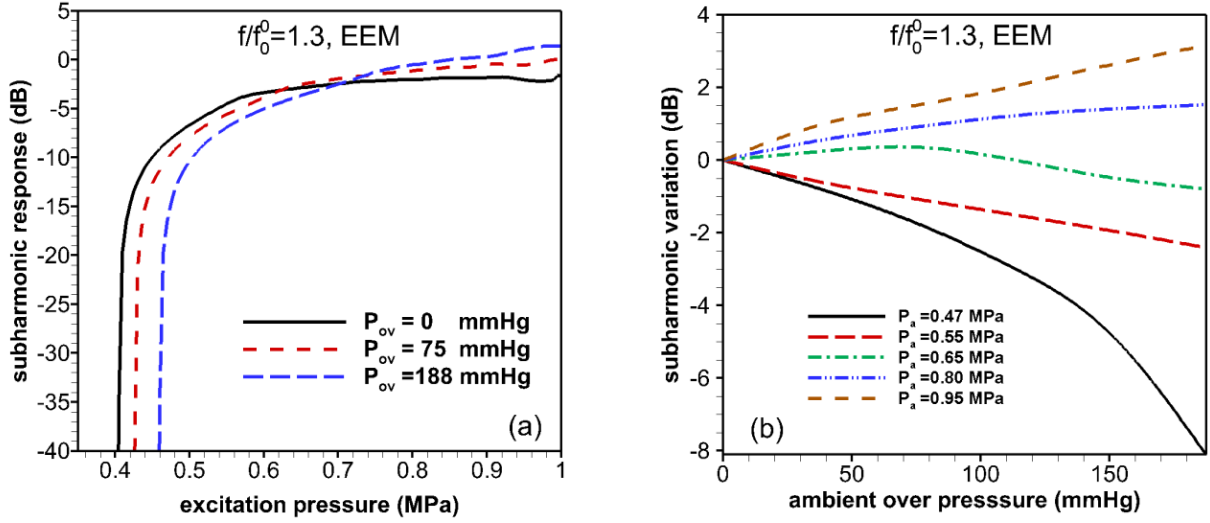

**Figure 6.** Subharmonic response of Sonazoid contrast agent varying with excitation pressure (a) and with ambient over pressure (b) at $\kappa^s$ = 3.0×10⁻⁸Ns/m with $R_0^0$ =3.0 μm according to EEM model.

**Effects of encapsulation damping and radius**

The difference between the behaviors of free microbubbles described in KS11 and encapsulated ones here primarily arises due to the fact that in contrast to free microbubbles, encapsulated microbubbles experience large shell damping, which significantly change the behavior. Most importantly, the range of excitations over which one gets subharmonic increases substantially. As was noted in our previous publication, damping due to encapsulation is the largest and increases with increasing surface dilatational viscosity $\kappa^s$ and decreasing radius (Katiyar and Sarkar, 2012). In order to investigate this effect, we consider two conditions at $f / f_0^0 = 1.3$. In Figure 5, we investigate the subharmonic response from a Sonazoid bubble of radius 1.6 μm, which is also the number average radius of this contrast agent. In Figure 6, we consider a Sonazoid microbubble of the radius 3 μm, but higher $\kappa^s$ = 1.2×10-8Ns/m. In both cases, due to the enhanced damping, the range of excitation pressures where one gets subharmonic response is wider; the curves cross at higher excitation. As a result, unlike in Figure 3, at the frequency ratio $f / f_0^0 = 1.3$, one obtains both decreasing and increasing subharmonic response with increasing overpressure depending on the excitation amplitude.

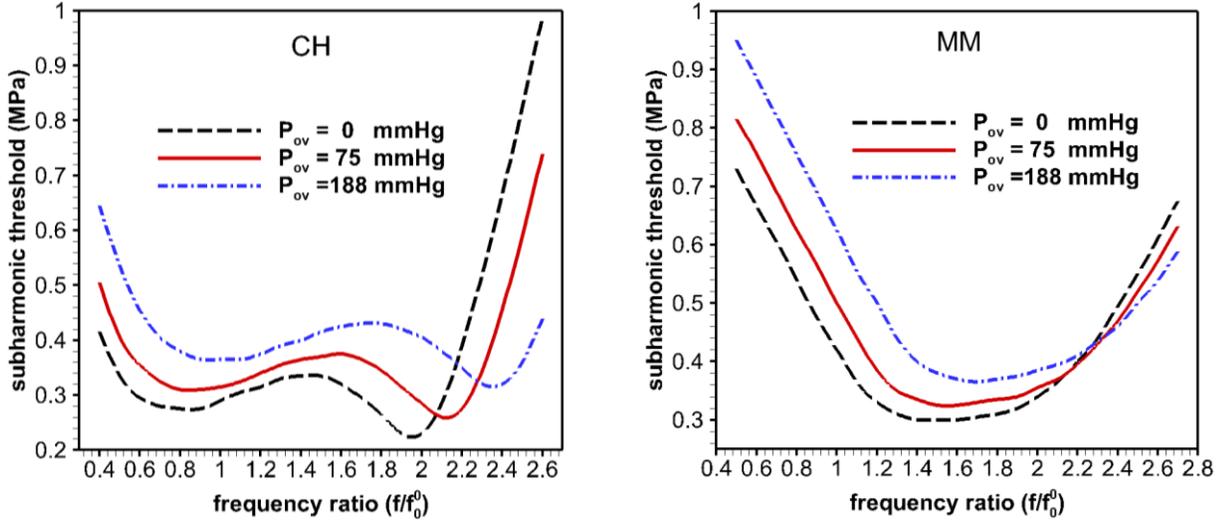

**Figure 7.** Subharmonic threshold according to Church-Hoff ( $f_0^0$ =1.8 MHz) and Marmottant ( $f_0^0$ =1.176 MHz) models for $R_0^0 = 3.0\,\mu m$.

## B. Effects of different rheological models

Figures 7 plots subharmonic thresholds as a function of frequency ratio for CH and MM models respectively. The curves are similar as in Figure 4 with different overpressures crossing around $f / f_0^0 \approx 2.0$. Correspondingly, we plot in Figures 8 and 9 subharmonic variation with overpressure at a lower and a higher frequency ratio.

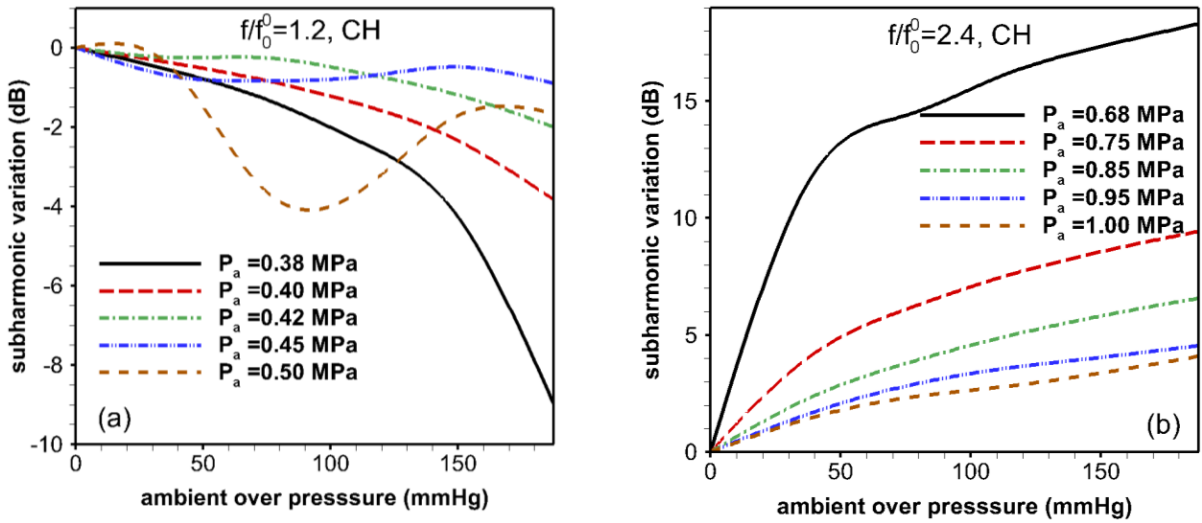

**Figure 8.** Subharmonic response of the contrast agent varying with ambient overpressure for (a) $f/f_0^0$ =1.2 and (b) =2.4 according to Church-Hoff model.

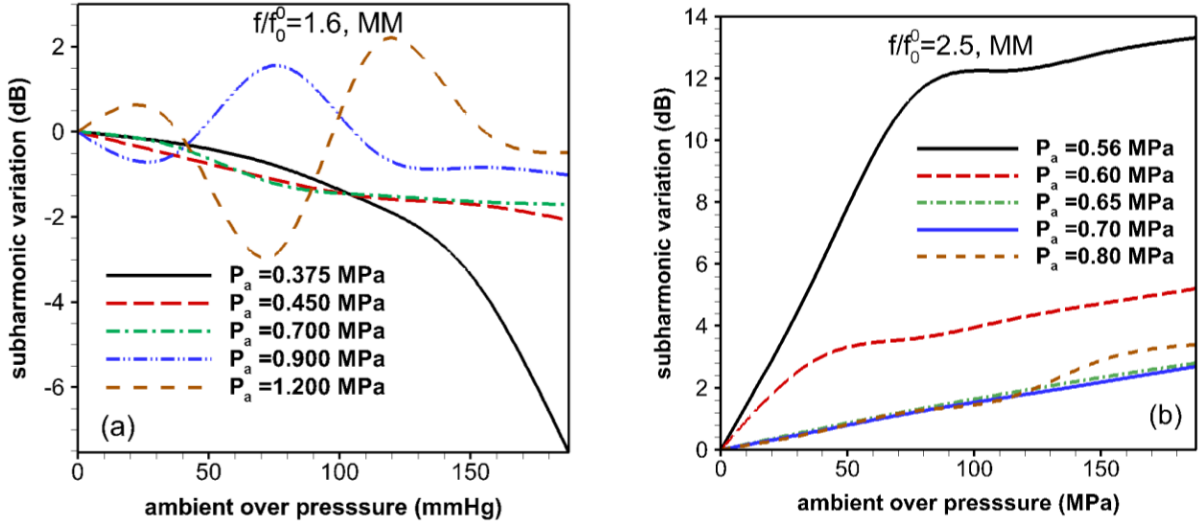

**Figure 9.** Subharmonic response of the contrast agent varying with ambient overpressure for (a) $f/f_0^0$ =1.2 and (b) =2.4 according to Church-Hoff model.

We notice that behaviors are similar to the EEM. At lower frequency ratios (Figure 8a and 9a), subharmonic response decreases or varies nonmonotonically with ambient overpressure, whereas, for higher frequency ratios $f/f_0^0 \approx 2.0$ it increases with increasing ambient overpressure. For lower excitation pressures the rise in subharmonic is steeper indicating an optimum range for subharmonic based pressure estimation.

## IV. Conclusion and Summary

We investigate subharmonic response of a contrast microbubble as a function of ambient overpressure. We use three different interfacial rheological models—strain softening exponential elasticity model (EEM), Church-Hoff model (CH) and Marmottant mdoel (MM). Different models produce qualitatively similar result. Therefore the Subharmonic variation with ambient pressure seen here is robust. It shows that the subharmonic response can either decrease, increase or vary nonmonotonically with increasing ambient pressure. There are important differences between behaviors of free and encapsulated microbubbles primarily because of the enhanced damping due to encapsulation which effectively increases the range of excitation for subharmonic response. For excitation frequency far higher than twice the resonance frequency of the microbubble, one obtains subharmonic increasing with overpressure, as was also the case for free microbubbles. But for lower excitation frequencies, one can obtain all possible trends depending on excitation pressure. The results indicate that one needs to carefully optimize the excitation frequency and amplitude for optimum performance and robustness of subharmonic emission based pressure estimation. Operating close to the excitation threshold, one can obtain higher sensitivity.

**Acknowledgement**

This work is partially supported by NSF Grants No. CBET-1205322, DMR-1239105 and GWU Katzen Cancer Research Center Innovative Cancer Pilot Research Award.

# References

Adam, D., Sapunar, M., and Burla, E. (**2005**). "On the relationship between encapsulated ultrasound contrast agent and pressure," Ultrasound Med. Biol. **31**, 673-686.

Andersen, K. S., and Jensen, J. A. (**2009**). "Ambient pressure sensitivity of microbubbles investigated through a parameter study," J. Acoust. Soc. Am. **126**, 3350-3358.

Andersen, K. S., and Jensen, J. A. (**2010**). "Impact of acoustic pressure on ambient pressure estimation using ultrasound contrast agent," Ultrasonics **50**, 294-299.

Bouakaz, A., Frinking, P. J. A., de Jong, N., and Bom, N. (**1999**). "Noninvasive measurement of the hydrostatic pressure in a fluid-filled cavity based on the disappearance time of micrometer-sized free gas bubbles," Ultrasound Med. Biol. **25**, 1407-1415.

Chang, P. H., Shung, K. K., Wu, S. J., and Levene, H. B. (**1995**). "Second-Harmonic Imaging and Harmonic Doppler Measurements with Albunex(R)," Ieee Transactions on Ultrasonics Ferroelectrics and Frequency Control **42**, 1020-1027.

Chatterjee, D., and Sarkar, K. (**2003**). "A Newtonian rheological model for the interface of microbubble contrast agents," Ultrasound Med. Biol. **29**, 1749-1757.

Church, C. C. (**1995**). "The Effects of an Elastic Solid-Surface Layer on the Radial Pulsations of Gas-Bubbles," J. Acoust. Soc. Am. **97**, 1510-1521.

Dave, J. K., Halldorsdottir, V. G., Eisenbrey, J. R., Merton, D. A., Liu, J. B., Zhou, J. H., Wang, H. K., Park, S., Dianis, S., Chalek, C. L., Lin, F., Thomenius, K. E., Brown, D. B., and Forsberg, F. (**2012a**). "Investigating the Efficacy of Subharmonic Aided Pressure Estimation for Portal Vein Pressures and Portal Hypertension Monitoring," Ultrasound Med. Biol. **38**, 1784-1798.

Dave, J. K., Halldorsdottir, V. G., Eisenbrey, J. R., Raichlen, J. S., Liu, J. B., McDonald, M. E., Dickie, K., Wang, S. M., Leung, C., and Forsberg, F. (**2012b**). "Subharmonic microbubble emissions for noninvasively tracking right ventricular pressures," American Journal of Physiology-Heart and Circulatory Physiology **303**, H126-H132.

deJong, N. (**1996**). "Improvements in ultrasound contrast agents," IEEE Engineering in Medicine and Biology Magazine **15**, 72-82.

deJong, N., Cornet, R., and Lancee, C. T. (**1994**). "Higher Harmonics of Vibrating Gas-Filled Microspheres .2. Measurements," Ultrasonics **32**, 455-459.

deJong, N., Hoff, L., Skotland, T., and Bom, N. (**1992**). "Absorption and Scatter of Encapsulated Gas Filled Microspheres - Theoretical Considerations and Some Measurements," Ultrasonics **30**, 95-103.

Fairbank, W. M., and Scully, M. O. (**1977**). "New Noninvasive Technique for Cardiac Pressure Measurement - Resonant Scattering of Ultrasound from Bubbles," IEEE Transactions on Biomedical Engineering **24**, 107-110.

Ferrara, K., Pollard, R., and Borden, M. (**2007**). "Ultrasound microbubble contrast agents: Fundamentals and application to gene and drug delivery," Annual Review of Biomedical Engineering **9**, 415-447.

Forsberg, F., Shi, W. T., and Goldberg, B. B. (**2000**). "Subharmonic imaging of contrast agents," Ultrasonics **38**, 93-98.

Frinking, P. J. A., Gaud, E., Brochot, J., and Arditi, M. (**2010**). "Subharmonic scattering of phospholipid-shell microbubbles at low acoustic pressure amplitudes," IEEE Transactions on Ultrasonics Ferroelectrics and Frequency Control **57**, 1762.

Goldberg, B. B., Raichlen, J. S., and Forsberg, F. (**2001**). *Ultrasound Contrast Agents: Basic Principles and Clinical Applications* (Martin Dunitz, London).

Halldorsdottir, V. G., Dave, J. K., Eisenbrey, J. R., Machado, P., Zhao, H., Liu, J. B., Merton, D. A., and Forsberg, F. (**2014**). "Subharmonic aided pressure estimation for monitoring interstitial fluid pressure in tumours - In vitro and in vivo proof of concept," Ultrasonics **54**, 1938-1944.

Halldorsdottir, V. G., Dave, J. K., Leodore, L. M., Eisenbrey, J. R., Park, S., Hall, A. L., Thomeniuv, K., and Forsberg, F. (**2011**). "Subharmonic Contrast Microbubble Signals for Noninvasive Pressure Estimation under Static and Dynamic Flow Conditions," Ultrason. Imaging **33**, 153-164.

Hoff, L., Sontum, P. C., and Hovem, J. M. (**2000**). "Oscillations of polymeric microbubbles: Effect of the encapsulating shell," J. Acoust. Soc. Am. **107**, 2272-2280.

Hok, B. (**1981**). "A New Approach to Non-Invasive Manometry - Interaction between Ultrasound and Bubbles," Medical & Biological Engineering & Computing **19**, 35-39.

Ishihara, K., Kitabatake, A., Tanouchi, J., Fujii, K., Uematsu, M., Yoshida, Y., Kamada, T., Tamura, T., Chihara, K., and Shirae, K. (**1988**). "New Approach to Noninvasive Manometry Based on Pressure Dependent Resonant Shift of Elastic Microcapsules in Ultrasonic Frequency-Characteristics," Japanese Journal of Applied Physics Part 1-Regular Papers Short Notes & Review Papers **27**, 125-127.

Itai, Y., and Matsui, O. (**1997**). "Blood flow and liver imaging," Radiology **202**, 306-314.

Kasimir, M. T., Seebacher, G., Jaksch, P., Winkler, G., Schmid, K., Marta, G. M., Simon, P., and Klepetko, W. (**2004**). "Reverse cardiac remodelling in patients with primary pulmonary hypertension after isolated lung transplantation," European Journal of Cardio-Thoracic Surgery **26**, 776-781.

Katiyar, A., and Sarkar, K. (**2010**). "Stability analysis of an encapsulated microbubble against gas diffusion," Journal of Colloid and Interface Science **343**, 42-47.

Katiyar, A., and Sarkar, K. (**2011**). "Excitation threshold for subharmonic generation from contrast microbubbles," J. Acoust. Soc. Am. **130**, 3137-3147.

Katiyar, A., and Sarkar, K. (**2012**). "Effects of encapsulation damping on the excitation threshold for subharmonic generation from contrast microbubbles," J. Acoust. Soc. Am. **132**, 3576-3585.

Katiyar, A., Sarkar, K., and Forsberg, F. (**2011**). "Modeling subharmonic response from contrast microbubbles as a function of ambient static pressure," J. Acoust. Soc. Am. **129**, 2325-2335.

Leodore, L., Forsberg, F., and Shi, W. T. (**2007**). "In Vitro Pressure Estimation Obtained from Subharmonic Contrast Microbubble Signals," IEEE Ultrasonics Symposium, 2207-2210.

Marino, T. A., Kent, R. L., Uboh, C. E., Fernandez, E., Thompson, E. W., and Cooper, G. (**1985**). "Structural-analysis of pressure versus volume overload hypertrophy of cat right ventricle," Am. J. Physiol. **249**, H371-H379.

Marmottant, P., van der Meer, S., Emmer, M., Versluis, M., de Jong, N., Hilgenfeldt, S., and Lohse, D. (**2005**). "A model for large amplitude oscillations of coated bubbles accounting for buckling and rupture," J. Acoust. Soc. Am. **118**, 3499-3505.

Nahire, R., Haldar, M. K., Paul, S., Ambre, A. H., Meghnani, V., Layek, B., Katti, K. S., Gange, K. N., Singh, J., Sarkar, K., and Mallik, S. (**2014a**). "Multifunctional polymersomes for cytosolic delivery of gemcitabine and doxorubicin to cancer cells," Biomaterials **35**, 6482-6497.

Nahire, R., Haldar, M. K., Paul, S., Mergoum, A., Ambre, A. H., Katti, K. S., Gange, K. N., Srivastava, D. K., Sarkar, K., and Mallik, S. (**2014b**). "Lipid nanopartciles with tunable echogenicity for targeted delivery to pancreatic cancer cells with simultaneous ultrasound imaging," Molecular Pharmaceutics **11**, 4059-4068.

Paul, S., Katiyar, A., Sarkar, K., Chatterjee, D., Shi, W. T., and Forsberg, F. (**2010**). "Material characterization of the encapsulation of an ultrasound contrast microbubble and its subharmonic response: Strain-softening interfacial elasticity model," J. Acoust. Soc. Am. **127**, 3846-3857.

Paul, S., Nahire, R., Mallik, S., and Sarkar, K. (**2014**). "Encapsulated microbubbles and echogenic liposomes for contrast ultrasound imaging and targeted drug delivery," Comput. Mech. **53**, 413-435.

Pieters, P. C., Miller, W. J., and DeMeo, J. H. (**1997**). "Evaluation of the portal venous system: Complementary roles of invasive and noninvasive imaging strategies," Radiographics **17**, 879-895.

Reddy, A. K., Taffet, G. E., Madala, S., Michael, L. H., Entman, M. L., and Hartley, C. J. (**2003**). "Noninvasive blood pressure measurement in mice using pulsed Doppler ultrasound," Ultrasound Med. Biol. **29**, 379-385.

Sarkar, K., Katiyar, A., and Jain, P. (**2009**). "Growth and dissolution of an encapsulated contrast microbubble " Ultrasound Med. Biol. **35**, 1385-1396.

Sarkar, K., Shi, W. T., Chatterjee, D., and Forsberg, F. (**2005**). "Characterization of ultrasound contrast microbubbles using in vitro experiments and viscous and viscoelastic interface models for encapsulation," J. Acoust. Soc. Am. **118**, 539-550.

Shankar, P. M., Krishna, P. D., and Newhouse, V. L. (**1998**). "Advantages of subharmonic over second harmonic backscatter for contrast-to-tissue echo enhancement," Ultrasound Med. Biol. **24**, 395-399.

Shankar, P. M., Krishna, P. D., and Newhouse, V. L. (**1999**). "Subharmonic backscattering from ultrasound contrast agents," J. Acoust. Soc. Am. **106**, 2104-2110.

Shi, W. T., Forsberg, F., Hall, A. L., Chia, R. Y., Liu, J. B., Miller, S., Thomenius, K. E., Wheatley, M. A., and Goldberg, B. B. (**1999a**). "Subharmonic imaging with microbubble contrast agents: Initial results," Ultrason. Imaging **21**, 79-94.

Shi, W. T., Forsberg, F., Liu, J., Needleman, L., and Goldberg, B. B. (**1999b**). "Nonlinear subharmonic imaging with US microbubble contrast agents," Radiology **213P**, 362-362.

Shi, W. T., Forsberg, F., Raichlen, J. S., Needleman, L., and Goldberg, B. B. (**1999c**). "Noninvasive pressure estimation with US microbubble contrast agents," Radiology **213P**, 101-101.

Shi, W. T., Forsberg, F., Raichlen, J. S., Needleman, L., and Goldberg, B. B. (**1999d**). "Pressure dependence of subharmonic signals from contrast microbubbles," Ultrasound Med. Biol. **25**, 275-283.

Strauss, A. L., Roth, F. J., and Rieger, H. (**1993**). "Noninvasive Assessment of Pressure - Gradients across Iliac Artery Stenoses - Duplex and Catheter Correlative Study," Journal of Ultrasound in Medicine **12**, 17-22.

**Figure Captions**

**Figure 1.** (a) Subharmonic response of Sonazoid microbubble ($R_0^0 = 3$ μm) as a function of $f / f_0^0$ ($f_0^0 = 1.76$ MHz). Subharmonic response as a function of ambient overpressure at excitation amplitude 0.34 MPa (b), 0.38 MPa (c) and 0.44 MPa (d) according to EEM model.

**Figure 2.** Subharmonic response of Sonazoid microbubble ($R_0^0 = 3$ μm) varying with excitation pressure (a) and with ambient over pressure (b) at $f / f_0^0 = 2.3$ according to EEM model.

**Figure 3.** Subharmonic response of Sonazoid contrast agent ($R_0^0 = 3$ μm) varying with excitation pressure (a) and with ambient over pressure (b) at $f / f_0^0 = 1.3$ according to EEM model.

**Figure 4.** Subharmonic threshold of Sonazoid contrast agent ($R_0^0 = 3$ μm) as a function frequency ratio according to EEM.

**Figure 5.** Subharmonic response of Sonazoid contrast agent varying with excitation pressure (a) and with ambient over pressure (b) at $f / f_0^0 = 1.3$ ($f_0^0 = 4.2$ MHz) with $R_0^0 = 1.6$ μm according to EEM model.

**Figure 6.** Subharmonic response of Sonazoid contrast agent varying with excitation pressure (a) and with ambient over pressure (b) at $\kappa^s = 3.0 \times 10^{-8}$ Ns/m with $R_0^0 = 3.0$ μm according to EEM model.

**Figure 7.** Subharmonic threshold according to Church-Hoff ($f_0^0 = 1.8$ MHz) and Marmottant ($f_0^0 = 1.176$ MHz) models for $R_0^0 = 3.0 \mu m$.

**Figure 8.** Subharmonic response of the contrast agent varying with ambient overpressure for (a) $f / f_0^0 = 1.2$ and (b) $= 2.4$ according to Church-Hoff model.

**Figure 9.** Subharmonic response of the contrast agent varying with ambient overpressure for (a) $f / f_0^0 = 1.2$ and (b) $= 2.4$ according to Church-Hoff model.